# Drinking water contamination as a population-wide determinant of mortality in California

Kiley Kennedy[1], Vincent Zheng[2], Benjamin Q Huynh[23*]

[1] Department of Epidemiology, Johns Hopkins Bloomberg School of Public Health
[2] Department of Environmental Health & Engineering, Johns Hopkins Bloomberg School of Public Health
[3] Institute for Planetary Health, Johns Hopkins University
[*] Corresponding author. Email: bhuynh@jhu.edu

**Abstract**

Drinking water contamination, a known determinant of adverse health outcomes, remains widespread and inequitably distributed amidst aging infrastructure. Regulatory oversight is the primary tool to protect drinking water-related public health risks, with maximum contaminant levels established to regulate concentrations for contaminants of concern. However, the extent to which existing concentrations of contaminants at the public water system level directly affect population health remains poorly understood. Here we present a large-scale data analysis of 20 million water samples from 2012-2022 in California to assess the impact of drinking water quality on all-cause mortality. Surfactants were associated with increases in mortality, potentially serving as proxies for wastewater contaminants. We further find evidence of mixture effects that were unidentifiable through single-contaminant analysis, suggesting mixtures of toxic metals as well as salinity constituents are associated with mortality. These results could inform public health efforts to mitigate mortality associated with consumption of contaminated drinking water.

**Introduction**

Drinking water contamination is a major public health concern. Contamination can be caused by sources such as industrial pollution, microbial contaminants, or naturally occurring geological sources, with a variety of associated adverse health outcomes. In particular, exposure to contaminated drinking water can cause infection, gastrointestinal distress, cancer, kidney disease, as well as developmental and neurological issues.[1,2]

The challenge of ensuring clean drinking water does not impact all communities equally. Social determinants such as racial and ethnic compositions, education level, age distributions, and socioeconomic status of populations have all been found to be associated with differential levels of contaminant concentrations, and decreased water quality.[3–10] For example, populations living in communities near farms are at an increased risk of receiving drinking water with industrial contaminants.[8,11]

In the United States, compliance with drinking water quality standards is largely based on the Environmental Protection Agency's (EPA) established Maximum Contaminant Levels (MCLs) and other requirements under the Safe Drinking Water Act. These levels are set for contaminants that are known to be present in the public water system and are proven to have adverse effects on health. The MCL set by the EPA is an enforceable standard and public water systems (PWS) that are unable to meet these standards are considered non-compliant or failing water systems.

Despite existing regulations, existing levels of contaminant concentrations could pose a substantial public health burden, even at low doses.[12] While links between specific drinking water contaminants and adverse health outcomes have been established at an individual level, it remains unclear how contamination at the level of a public water system translates into changes in population-wide health outcomes, such as mortality. Quantifying the mortality burden attributable to population-wide water contaminants could inform regulatory updates and targeted interventions towards public water systems found to be most adversely affecting population health. Similar to air pollution research, which often examines population-wide health outcomes based on large-scale exposures, drinking water quality may also have a measurable impact on population-level mortality.

Data collected as part of regulatory compliance with the Safe Drinking Water Act provide a unique opportunity to assess the relationship between drinking water quality and population-wide mortality. Here we conduct a large-scale data analysis of 20 million water samples from 2012-2022 in California, assessing the impact of 55 different water contaminants on all-cause mortality. We do so by conducting a water-wide association analysis screening all analytes, followed by robustness checks. We further identify co-occurring contaminants in mixtures and estimate their aggregate effects on mortality. Our results offer potential regulatory targets for investigation to reduce mortality burden arising from drinking water contaminants.

**Results**

*Drinking water analytes associated with mortality*

Surfactants were associated with increased mortality, with each standard deviation increase in concentration associated with 0.46% (95% Confidence Interval: [0.14, 0.79]) increases in all-cause mortality (Figure 1). This estimated association corresponds to 5650 (95% CI: [2296, 10619]) deaths from 2012-2022 in terms of attributable mortality, or 513 deaths per year (Figure 2). Dibromoacetic acid was associated with reduced mortality (-0.66%, [-1.07, -0.24]). (Supplementary Table 1)

Mortality increases associated with analytes may be proxy signals for treatment and source-water regimes rather than direct toxic effects of the analytes themselves. Surfactants, arising from both household detergents and industrial waste, may have an association with mortality as a proxy for wastewater contaminants. Similarly, dibromoacetic acid, a disinfection byproduct, may have a negative association with mortality arising from the protective effect of disinfectants outweighing the harms of disinfection byproducts.

Among 55 analytes, nine were associated with mortality after multiple-testing correction; only the two noted above remained robust across alternative specifications and sensitivity analyses (Figure 1). Associations for Radium-228, 1,2-Dibromo-3-Chloropropane, Chloroform, Uranium, Total Carbon, Carbonate Alkalinity, and Perchlorate did not withstand these checks.

*Distributed lag and exposure-response analyses*

Additional analyses incorporating lagged exposure values were incorporated to test for cumulative effects (Figure 3). After including 1- and 2-year lags, the cumulative effect of surfactants was a 1.87% (0.71, 3.05) increase in mortality per standard deviation. Dibromoacetic acid had a cumulative effect of -1.39% (-2.29, -0.49). Exposure-response analysis indicates the identified associations for surfactants and dibromoacetic acid are largely linear, particularly at values at or below the 99th percentile (Figure 4).

Radium-228 and Chloroform – also a disinfection byproduct – had large cumulative effects of 1.74% (0.96, 2.54) and -1.77% (-2.87, -0.66). However, we incorporated a leading variable as a negative control, and found non-null associations for the leading variable, suggesting potential residual confounding for these two analytes (Figure 3).

*Joint impacts from mixtures of drinking water contaminants*

We identified three distinct clusters of co-occurring drinking water analytes across all 55 analytes tested. We used three different cluster visualization methods – a pairwise correlation heatmap, a correlation network, and multidimensional scaling – and found similar results across all three methods (Figure 5). The three clusters identified can broadly be defined as disinfection byproducts, industrial pollutants, and salinity ions.

Clusters of disinfection byproducts included analytes such as dibromoacetic acid, bromoform, dibromochloromethane, bromodichloromethane, total trihalomethanes, trichloroacetic acid, chloroform, total haloacetic acids, and dichloroacetic acid. Clusters of industrial pollutants largely centered around mercury, cyanide, and perchlorate. Clusters of salinity ions or salinity-related metrics centered around sodium, chloride, and sulfate.

We used these identified clusters of co-occurring contaminants to estimate mixture effects on mortality. We also included another mixture of toxic metals including lead, arsenic, mercury, chromium, barium, copper, manganese, and aluminum, based on prior evidence indicating potential adverse health outcomes.

A quartile increase in the toxic metals mixture was associated with a 6.77% (2.32,11.42) increase in mortality. Effect sizes for disinfection byproducts, industrial contaminants, and salinity ions were -1.03% (-1.86,-0.19), 5.63% (-1.16,12.89), and 2.17% (1.41,2.95), respectively (Figure 6).

**Discussion**

Our analysis indicates that some key drinking water analytes, such as surfactants, toxic metals, and salinity ions, are associated with all-cause mortality. As with any observational study, residual confounding is possible. However, by using two-way fixed effects with zip-code and year indicators and adjusting for income, age structure, and water source, we effectively adjust for time-invariant place factors and common temporal shocks; any remaining confounder must therefore vary within zip codes over time and co-vary with changes in water chemistry. Such co-variation is most plausibly driven by other aspects of drinking-water quality (e.g., other unobserved contaminants or interactions between analytes). Therefore, even if our identified associations represent proxies rather than causal toxicants, the signals we detect are still likely to arise from drinking water-related risk. Taken together, our results indicate that widespread, low-dose exposure to drinking water contaminants may increase population-level mortality, warranting further investigation into underlying mechanisms.

In particular, the detected associations for surfactants and disinfection byproducts are likely proxies rather than causal effects. Levels of surfactants, also known as methylene blue active substances, are indicative of detergents or foaming agents in water supplies. The presence of surfactants in drinking water may therefore indicate contamination from industrial or residential wastewater, which could plausibly impact mortality via complex contaminant mixtures. That the cumulative effect aggregated over three years was larger than the direct, contemporaneous effect of surfactant exposure suggests the impact on mortality may accumulate over time.

By similar logic, the observed reduction in mortality associated with disinfection byproducts (DBPs) probably comes from the protective effect of disinfectant use rather than the impact of DBPs themselves. DBPs are known to have deleterious effects on human health, but disinfectants are still widely used as their benefits are considered to outweigh the potential harms. Because water samples are taken at households and therefore downstream of water treatment, identifying the specific effect of DBPs on mortality is implausible given our available data, as constructing the relevant counterfactual would require adjusting for pre-treatment microbial levels. We therefore interpret the inverse associations between DBPs and mortality as net protection from disinfection use, not as evidence towards the health impacts of DBPs.

Our mixture analyses identified toxic metals and salinity ions as associated with mortality despite their individual analytes not being flagged by our initial water-wide association screening. This potentially reflects small, correlated, and interactive mixture effects not captured by single-exposure analysis. The toxic metals mixture could plausibly have a pathway towards mortality risk via cardiovascular or kidney diseases, or represent an index of corrosion and metal exposure more generally. Similarly, salinity ions could plausibly affect mortality via increasing sodium load for salt-sensitive populations, increased corrosion, or general chemistry shifts that could impact co-occurring contaminants.

Radium-228 was associated with increased mortality, although its non-null effect for the leading variable negative control indicates it may be subject to residual confounding. This does not necessarily mean that radium-228 has a null association with mortality, but more likely that its true effect relative to what we observe is attenuated by confounders. The estimated effect size for radium-228 is implausibly large, corresponding to over 2,000 attributable deaths annually. We hypothesize its true effect on mortality is much smaller, and that the estimated effect may have been confounded by co-occurring groundwater contaminants. Nevertheless, all values of radium-228 in our analysis were far below the federally regulated maximum contaminant level of 5 pCi/L; if a causal effect specific to radium-228 exists, it would have substantial public health implications.

Our estimated exposure-response curves appeared to be linear, especially at concentration levels at or below the 99th percentile. We therefore see no visual nonlinearities indicating "safe" levels of exposure; marginal effects appeared constant rather than accelerating only at higher concentrations. This is crucial for regulatory efforts, as even small doses of exposures applied to entire populations, as drinking water contaminants are, could have outsize impacts on population health that are imperceptible at the individual level.[12]

Our work has several limitations. First, zip codes with fewer than 11 annual deaths had their mortality counts censored to 0 for privacy purposes, which may have biased our estimates in low-population areas. This constituted 15.6% of observations, and prevented us from using more specific mortality metrics such as age-specific or cause-specific mortality, which would be further subject to censoring. Furthermore, because all-cause mortality is a distal, multi-factorial outcome, any identified effect sizes will be relatively modest, considering the myriad contributors to mortality. The high-dimensional nature of the dataset limits our statistical power to make robust inferences, particularly when expecting modest effect sizes. Our approach to adjust for unit- and time- fixed effects while also adjusting for income, age structure, and water source served as a general approach to adjusting for confounders in a water-wide association screening, but we did not employ analyte-specific covariate adjustments post-hoc in order to conserve statistical power. Similarly, many analytes were subject to limits of detection, requiring imputation that may have biased exposure estimates. Additionally, this study employs an ecological regression analysis, comparing population-level exposures to population-level health metrics, limiting the ability to draw conclusions about individual-level risk.

Our work draws on previous work that has established the evidentiary basis for drinking water as a determinant of health. In particular, prior studies include a water-wide association study to identify the effect of contaminants in groundwater wells on kidney cancer,[13] as well as broader studies involving environment-wide association studies.[14,15] Relevant previous work also includes California-specific drinking water studies,[16,17] and epidemiological studies assessing the impact of various contaminants on mortality.[18–25] Our work contributes to the literature by conducting a state-wide and water-wide analysis over 55 analytes, identifying single-analyte associations with mortality as well as mixture effects.

In summary, this study provides insights into the potential impacts of various drinking water analytes on all-cause mortality. A water-wide screening approach paired with mixture analyses may identify associations that either approach alone may miss. Given the potential for widespread population-level health impacts from even low-level environmental exposures, these findings warrant improved drinking water data collection to pinpoint mechanisms underlying the effect of drinking water contaminants on mortality.

**Methods**

This cross-sectional study was a secondary data analysis of deidentified and delinked publicly available data sources. It did not require review based on the Institutional Review Board process at Johns Hopkins University. To perform this cross-sectional analysis, we combined multiple secondary data sources to identify locations of PWS, estimate the drinking water quality of each zip-code, and estimate the health of each community through all cause mortality rates.

*Data Sources*

We used publicly available data from the California State Water Resources Control Board, and the Division of Drinking Water (DDW), the agency responsible for regulating 7,500 PWS throughout the State of California.[26] This regulatory oversight requires the reporting of water-quality sampling to DDW. Between the years 2011 and 2022 19,586,330 samples were collected and reported to the DDW. These samples represent all 7,500 PWS the DDW is responsible for overseeing, with 257 unique analytes being reported. We obtained the California Drinking Water System Area Boundaries, as well as the California Zip-Code Shapefile from the California State Geoportal.[27,28] These files allowed us to map each PWS in the state to its corresponding zip-code, allowing for the geographic-area for each PWS to be calculated. Population counts and the median household income (in 2023 inflation-adjusted dollars) for each zip-code were obtained using the 5-year estimates from the American Community Survey,[29] and death profiles by zip code were accessed from the California Department of Public Health.[30]

*Analyte Concentrations*

Analyte concentrations below the limit of detection (LOD), recorded as 0.00, were imputed using LOD divided by the square root of 2. Given that multiple public water systems (PWS) may serve a single zip-code and some PWS may span multiple zip-codes, a spatial mapping approach was employed. Using the WGS 84 coordinate system, each PWS was assigned a geographic area, enabling the calculation of zip-code-level analyte concentrations. For each zip-code, the weighted median analyte concentration was computed across all PWS within that region. This process yielded a dataset where each zip-code had a single record per year (over 11 years), containing the median concentration values for all analytes. Analytes that were not sampled in a given year within a zip-code were considered missing and assigned a value of NA.

*Primary Water Source*

Primary water source information was obtained directly from the California State Water Resources Control Board, and the Division of Drinking Water (DDW). Water sources were classified into 6 categories: ground water (GW; n = 6415), ground water paid (GWP; n = 62), ground water under influence of surface water (GU; n = 111), surface water (SW; n = 621), surface water paid (SWP; n = 434), and private (NA; n = 172). The six water source categories we collapsed into a binary variable indicating groundwater use. PWS utilizing either GW or GWP were designated as groundwater while all other PWS were categorized as non-ground water facilities. For zip-codes with multiple PWS presence of any ground water within the zip-code resulted in classification as groundwater.

*Percentage of Population in Age Categories*

Age categories of interest were selected based on biologically relevant exposure windows. These categories were determined to be: <5 years, 5–14 years, 15–24 years, 25–64 years, and ≥65 years. For each zip-code and year the percentage of the population within each age group was calculated by dividing the number of individuals in the age category by the total zip-code population for that corresponding year and multiplying by 100.

*Analytic Approach*

Analytes with ≥50% missing values were excluded from the dataset, along with those exhibiting a near-zero variance. Near-zero variance was defined as a variable with a frequency ratio greater than 19 (95/5) and a percentage of unique values less than 0.1024%. The remaining analytes were *z*-score standardized to facilitate comparability across different scales of sampling units and effect estimates, resulting in a final dataset containing 55 analytes, standard errors were clustered by zip-code to account for the fact that analyte concentrations within each zip-code are not independently and identically distributed.[31]

To estimate the association between water analyte concentrations and mortality rates, a two-way fixed effects regression model was employed. Each analyte was modeled with zip-code and year included as fixed effects to adjust for time-invariant and temporal confounders. Additionally, each model was adjusted for median household income, primary water source, and percentage of population in each age category (<5 years, 5–14 years, 15–24 years, 25–64 years, and ≥65 years). The model was specified as a Poisson regression modeling mortality rate, with the raw all cause mortality counts as the outcome and the log of the total population included as the offset term:

$$y_{it}|A_{it}, X_{it}, \alpha_i, \delta_t \sim Poisson(\lambda_{it}), \; log\lambda_{it} = logN_{it} + \beta A_{it} + X'_{it}\gamma + \alpha_i + \delta_t$$

where $y_{it}$ is the number of deaths in zip code $i$ and year $t$; $A_{it}$ is a given analyte's concentration value; $X_{it}$ is the matrix of adjustment covariates including median household, percentage of population in age categories, and a binary indicator for whether groundwater was the primary water source; and $\alpha_i$ and $\delta_t$ are the zip code and year fixed effects, respectively. Using the fixest package in R allowed for estimation of fixed-effects to be done via a within-transformation, which avoids estimating a coefficient for each fixed-effect level. The within transformation utilizes an alternating projection method, based on the iterative approach using a demeaning algorithm presented by Guimarães and Portugal in 2010.[32] A Benjamini-Hochberg (BH) adjustment was applied to adjust the p-value, reducing the probability of false positives.[33]

Robustness checks included alternative specifications to our regression model: (1) using age-adjusted mortality as the outcome variable instead of adjusting for age-structure, (2) using no covariates for adjustment, (3) using only median income and age structure as adjustment covariates, and (4) only adjusting for age structure.

*Secondary Analysis*

To further characterize the relationship between water contaminant exposure and mortality, several secondary analyses were performed. These included: correlation and co-occurrence analysis of water contaminants, quantile g-computation (qgcomp) to estimate joint mixture effects, distributed lag models, and evaluation of nonlinear exposure-response functional forms using generalized additive models (GAMs).

*Correlation and Co-occurrence Analysis*

To evaluate patterns of co-occurrence among contaminants we calculated pairwise Pearson correlation coefficients across all analytes. These correlations were then visualized using both a heatmap and a network plot restricted to |r| > 0.3. A dissimilarity matrix was constructed using the absolute Pearson correlation coefficients (distance = 1 - |r|). Highly correlated analytes, regardless of direction, were considered similar. Classical multidimensional scaling (MDS) was

applied to visualize analytes in a two-dimensional space based on their patterns of co-occurrence.[34] Their chemical class was then added as a layer to the plot to facilitate interpretation of clustering patterns by contaminant type, highlighting class specific groupings.

*Quantile Based G-Computation*

To investigate potential joint effects of contaminants inside these co-occurring clusters on mortality, an approach known as quantile-based g-computation was utilized.[35] Contaminants were grouped into mixtures based on their empirical correlation structure, derived from pairwise Pearson correlation across all samples and further informed using knowledge of known chemical clusters. Four mixtures were defined; toxic metals, industrial contaminants, inorganic ions, and disinfection byproducts. The toxic metal mixture included: lead, arsenic, mercury, chromium hex (hexavalent chromium), barium, copper, manganese, and aluminum. The industrial contaminants mixture included: perchlorate, mercury, cyanide. The inorganic ions mixture included: chloride, sodium, total dissolved solids (TDS), sulfate, conductivity, and hardness total as CaCO3. The disinfection byproduct mixture included: bromodichloromethane, bromoform, chloroform, dibromoacetic acid, dibromochloromethane, dichloroacetic acid, total haloacetic acids HAA5, trihalomethanes (TTHM), trichloroacetic acid, and ethylene dibromide. An adaptation of weighted quantile sum regression, quantile g-computation allows for heterogeneous directionality of estimates within the mixture. Each mixture was modeled as a single quantized index formed by summing the ranked (quantile-transformed) exposure variables. Poisson regression models were then used to estimate the overall effect of increasing the mixture index by one quartile, adjusting for the same covariates as in the primary analysis. However, unlike in the primary analysis year was treated as a basis spline rather than a categorical factor, modeling nonlinear trends over time rather than estimating separate effects for each year.

*Distributed lag model*

To investigate potential delayed effects of exposure, we extended the Poisson regression framework described above to include lagged exposure variables via a distributed lag model. Specifically, exposure variables were lagged by 1 and 2 years, while all other covariates and fixed effects remained aligned with the outcome year. This approach allowed us to assess whether contaminant levels in prior years were associated with mortality in subsequent years, capturing potential time-lagged associations between exposure and health outcomes. We also incorporated a lead variable by 1 year as a negative control.

*Generalized additive models*

Nonlinear functional forms between contaminant concentrations and mortality were explored by fitting generalized additive models using penalized smoothing splines. The models were adjusted for the same covariates and fixed effects as the main Poisson models. The slope of the exposure-response curve was visualized by calculating and plotting the first derivative of the fitted spline on the response scale.

*Data availability*

All data used in this study are publicly available. Code for analyses will be made publicly available upon acceptance for publication as a peer-reviewed manuscript.

*Author contributions*

KK: methodology, formal analysis, and writing of original draft. VZ: preliminary analysis. BQH: conceptualization, methodology, analysis, supervision, and project administration. All authors contributed to the writing of the final manuscript.

*Competing interests*

Authors declare no competing interests.

# Figures

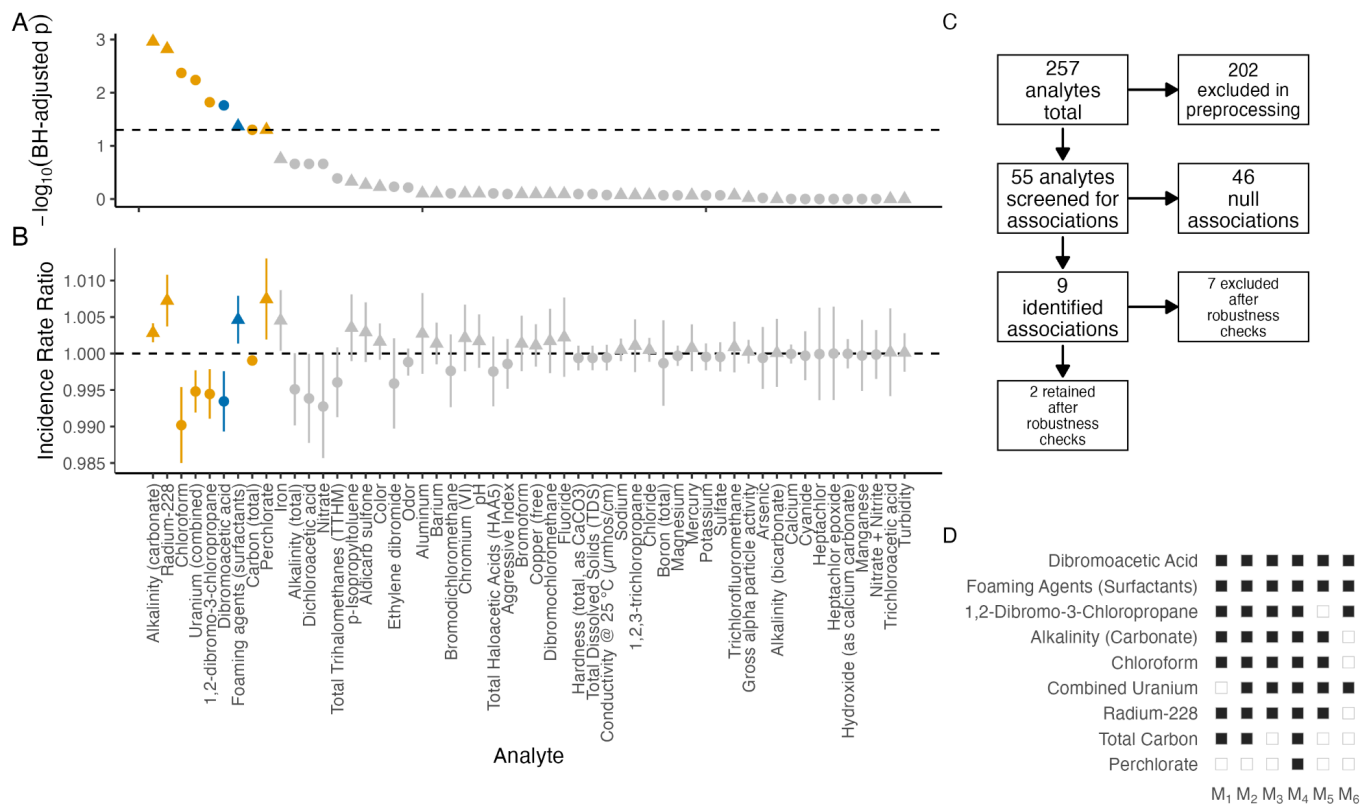

**Figure 1: Drinking water analytes screened for associations with all-cause mortality.** Panel A depicts the negative log of *p*-values adjusted for multiple hypothesis testing via the Benjamini-Hochberg method. Panel B depicts the incidence rate ratio, or the increase in mortality rate per each standard deviation increase in concentration per analyte. Triangles indicate a positive association; circles indicate a negative association. Vertical lines indicate 95% confidence intervals. Blue indicates an association retained after robustness checks; orange indicates those excluded after checks. Gray shapes are associations considered not statistically significant. Panel C depicts the screening and robustness check process. Panel D depicts which of the screened analytes passed each of the robustness checks. $M_1$ used a different outcome variable specification; $M_2$, $M_3$, and $M_4$ used different covariate adjustment schemes; $M_5$ checked for cumulative effects in a distributed lag model (DLM); $M_6$ used a leading term in the DLM as a negative control.

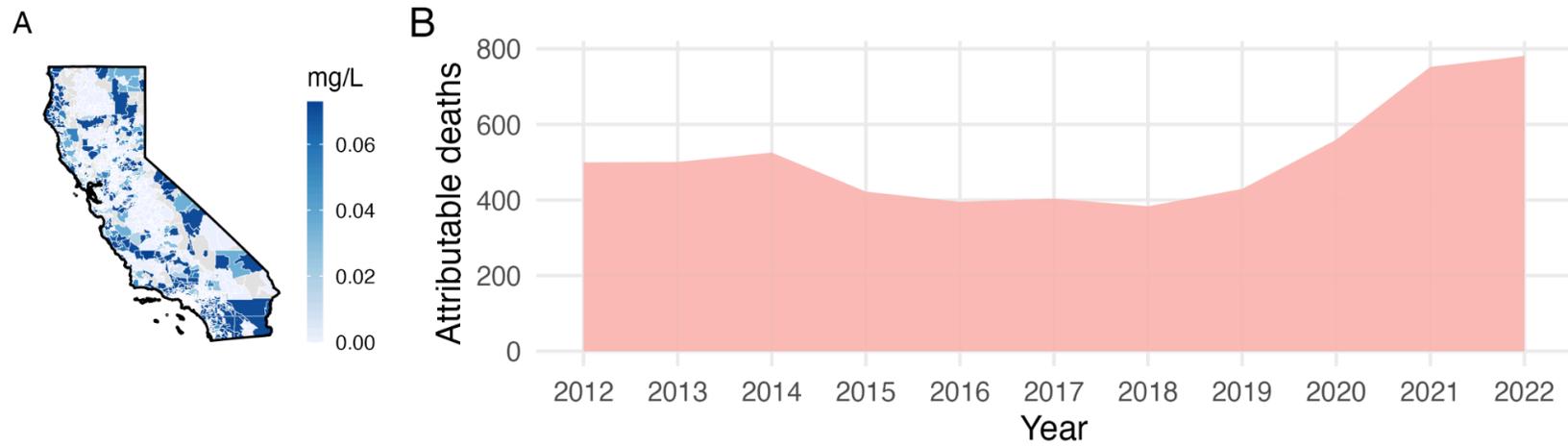

**Figure 2: Geographical concentrations and estimated attributable mortality for surfactants.** Panel A depicts zip code-level median concentrations of surfactants from 2012-2022. Panel B depicts estimated mortality associated with surfactants.

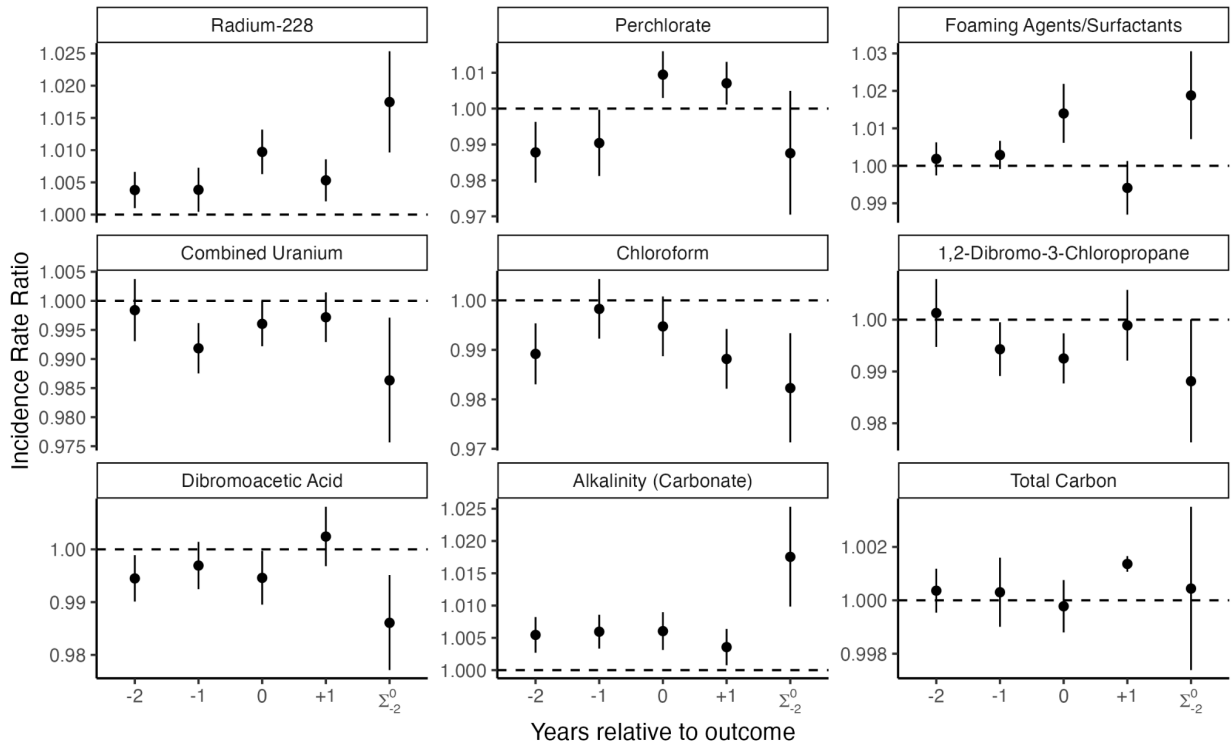

**Figure 3: Distributed lag model effect sizes for screened drinking water analytes.** Σ indicates the cumulative effect from summing the effect sizes from -2 years to 0 years. Vertical lines indicate 95% confidence intervals.

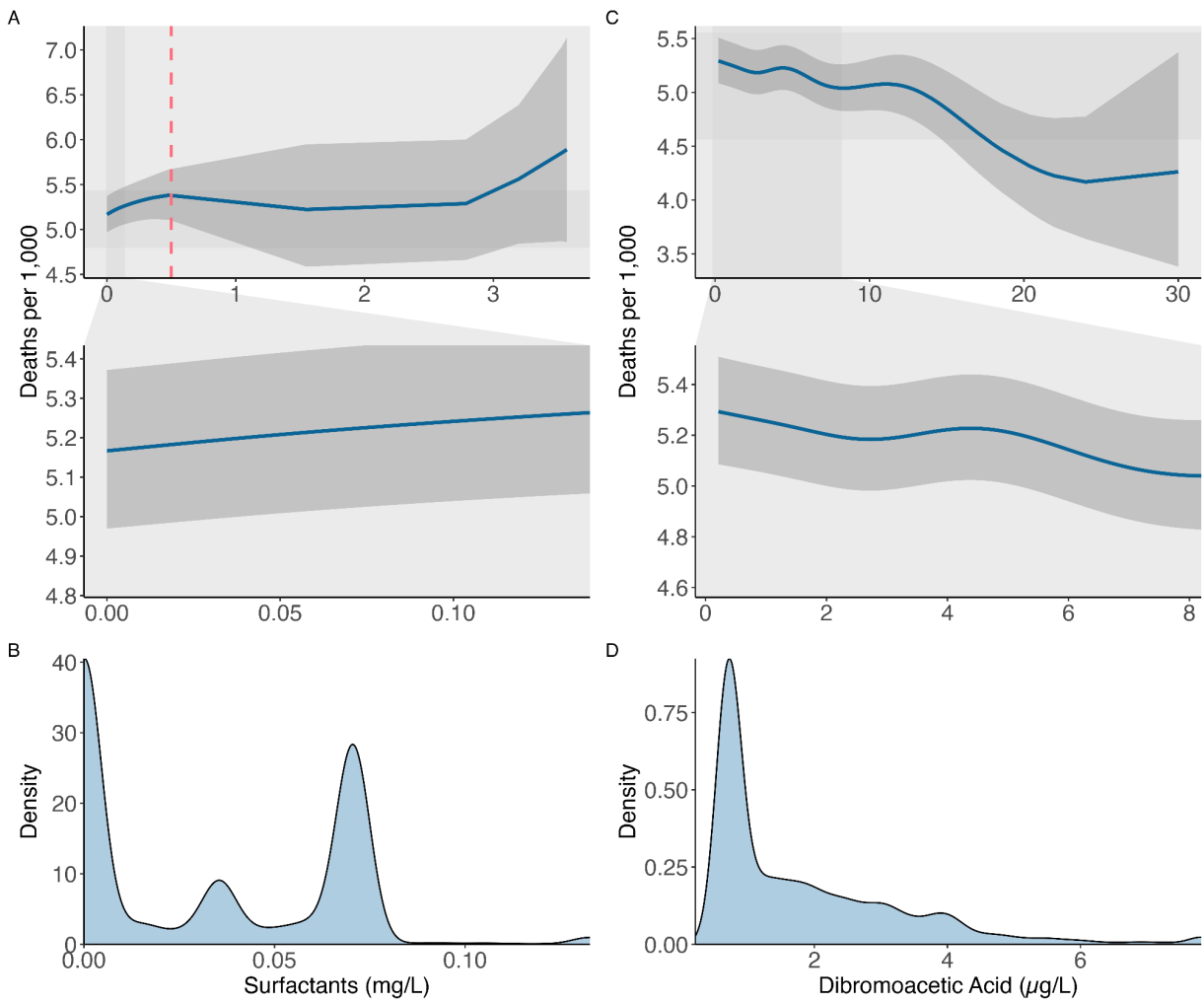

**Figure 4: Estimated exposure-response curves for Surfactants and Dibromoacetic Acid.** Panels A and C depict exposure-response curves estimated by generalized additive models, with zoomed-in subpanels indicating values up to the 99th percentile of concentration values. Red dashed line indicates the regulatory maximum contaminant level for an analyte. Panels B and D are density plots of concentration values for each analyte, with values above the 99th percentile top-coded for visualization purposes.

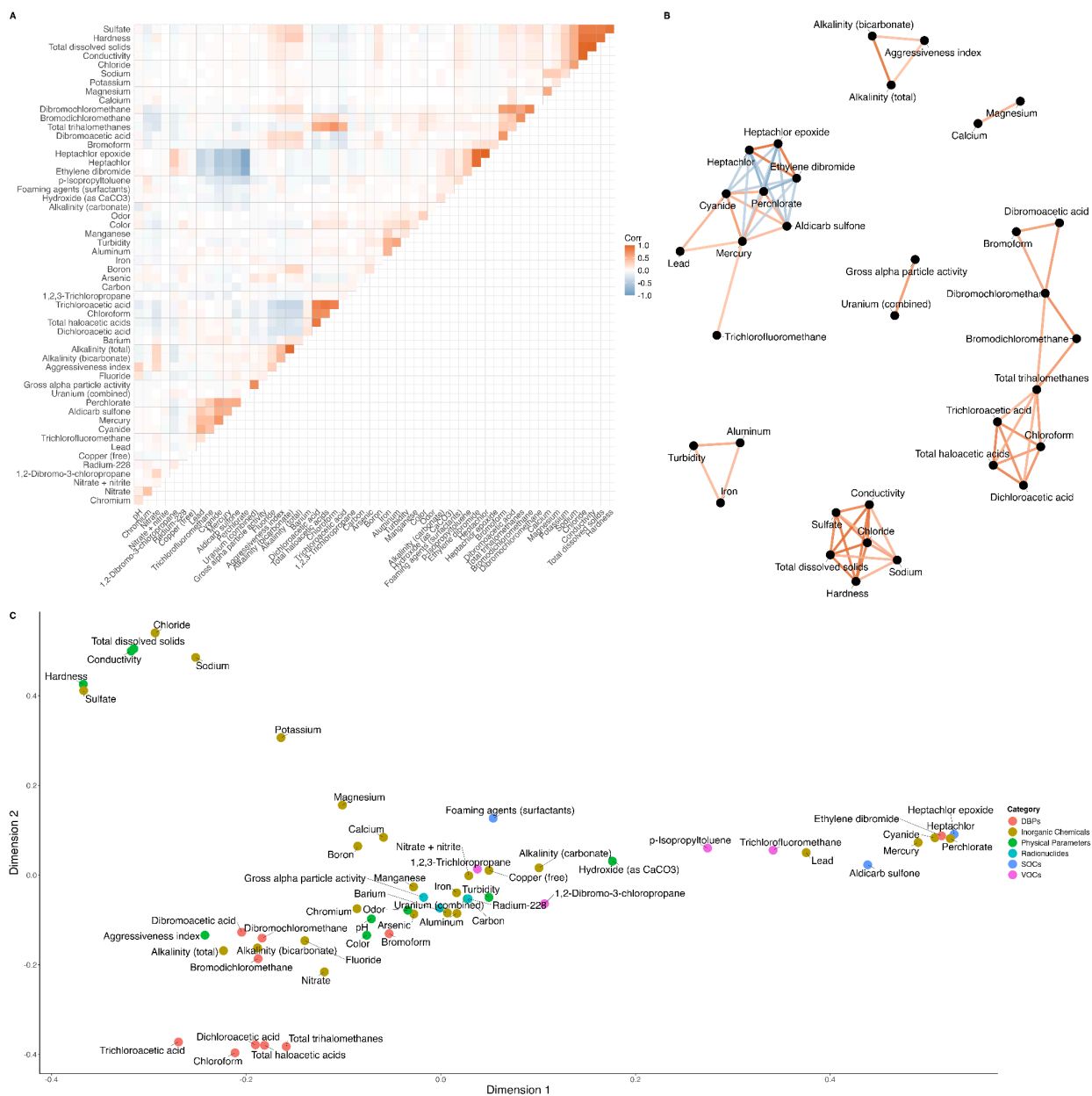

**Figure 5: Correlation structure of drinking water analytes.** Panel A depicts a pairwise correlation heatmap. Panel B depicts a correlation graph network. Panel C depicts a multi-dimensional scaling applied to a distance matrix of all analytes, where clustered points indicate similar correlation profiles.

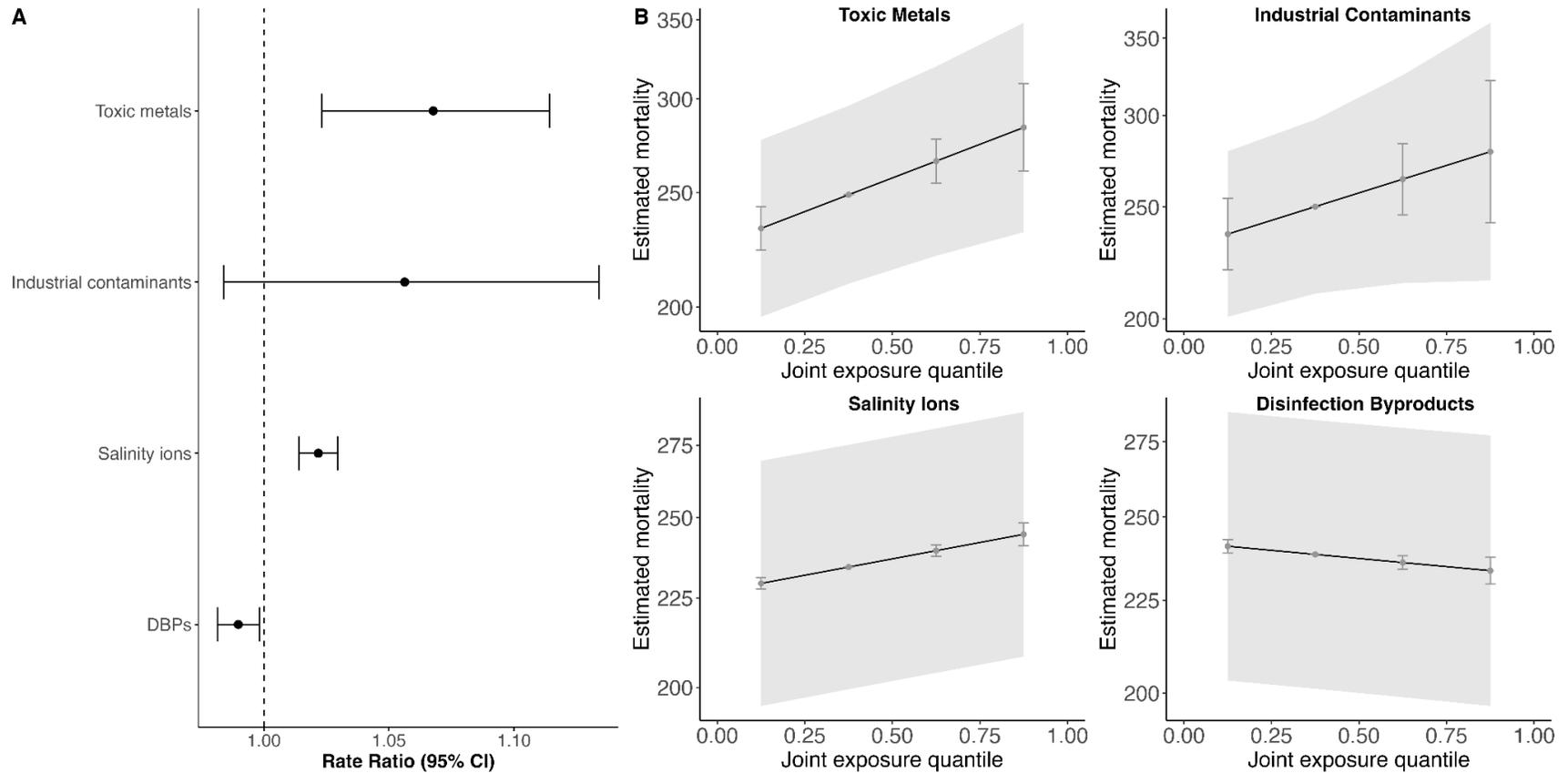

**Figure 6: Mixture effects identified by quantile g-computation.** Panel A depicts the effect sizes of drinking water contaminant mixtures, and Panel B depicts their exposure-response relationships quantified by increasing all components of a mixture by one quantile. Toxic Metals refers to a mixture of lead, arsenic, mercury, chromium, barium, copper, manganese, and aluminum. Industrial Contaminants refers to a mixture of perchlorate, cyanide, and mercury. Salinity Ions refers to a mixture of sodium, chloride, and sulfate. Disinfection Byproducts refers to a mixture of bromidichloromethane, bromoform, chloroform, dibromoacetic acid, dibromochloromethane, dichloroacetic acid, trichloroacetic acid, ethylene dibromide, total haloacetic acids, and total trihalomethanes.

# Supplemental Material

**Table 1: Regression results for the water-wide association screening.** BH-p denotes p-value after Benjamini-Hochberg adjustment.

| Analyte | Coefficient | Std. Err | Increase (95% CI) | *p*-Value | BH-p |
|---|---|---|---|---|---|
| **Disinfection byproducts** | | | | | |
| Chloroform | -0.0099 | 0.003 | -0.98 (-1.50, -0.46) | <0.001 | 0.004 |
| Dibromoacetic Acid | -0.0066 | 0.002 | -0.66 (-1.07, -0.24) | 0.002 | 0.017 |
| Dichloroacetic Acid | -0.0062 | 0.003 | -0.62 (-1.22, -0.00) | 0.049 | 0.218 |
| Trihalomethanes (TTHM) | -0.004 | 0.002 | -0.40 (-0.87, 0.08) | 0.106 | 0.409 |
| Ethylene Dibromide | -0.0041 | 0.003 | -0.41 (-1.03, 0.21) | 0.193 | 0.588 |
| Bromodichloromethane | -0.0024 | 0.003 | -0.24 (-0.74, 0.26) | 0.35 | 0.784 |
| Total Haloacetic Acids (HAA5) | -0.0025 | 0.002 | -0.25 (-0.72, 0.23) | 0.313 | 0.784 |
| Bromoform | 0.0014 | 0.002 | 0.14 (-0.24, 0.52) | 0.477 | 0.807 |
| Dibromochloromethane | 0.0017 | 0.002 | 0.17 (-0.27, 0.61) | 0.446 | 0.807 |
| Trichloroacetic Acid | 0.0002 | 0.003 | 0.02 (-0.58, 0.62) | 0.959 | 0.999 |
| **Inorganic Chemicals** | | | | | |

| Analyte | Coefficient | Std. Err | Increase (95% CI) | p-Value | BH-p |
|---|---|---|---|---|---|
| Carbonate Alkalinity | 0.0028 | 0.001 | 0.28 (0.15, 0.41) | <0.001 | 0.001 |
| Perchlorate | 0.0074 | 0.003 | 0.75 (0.19, 1.30) | 0.008 | 0.05 |
| Iron | 0.0045 | 0.002 | 0.45 (0.04, 0.87) | 0.033 | 0.177 |
| Total Alkalinity | -0.0049 | 0.003 | -0.49 (-0.98, 0.00) | 0.053 | 0.218 |
| Nitrate | -0.0073 | 0.004 | -0.73 (-1.43, -0.01) | 0.045 | 0.218 |
| Aluminum | 0.0027 | 0.003 | 0.27 (-0.28, 0.83) | 0.331 | 0.784 |
| Barium | 0.0014 | 0.001 | 0.14 (-0.15, 0.42) | 0.347 | 0.784 |
| Chromium Hex | 0.0021 | 0.002 | 0.21 (-0.24, 0.67) | 0.359 | 0.784 |
| Free Copper | 0.0011 | 0.001 | 0.11 (-0.18, 0.40) | 0.452 | 0.807 |
| Fluoride | 0.0022 | 0.003 | 0.22 (-0.32, 0.77) | 0.421 | 0.807 |
| Sodium | 0.0005 | 0.001 | 0.05 (-0.10, 0.20) | 0.517 | 0.84 |
| Chloride | 0.0005 | 0.001 | 0.05 (-0.11, 0.21) | 0.549 | 0.847 |
| Total Boron | -0.0013 | 0.003 | -0.13 (-0.72, 0.45) | 0.656 | 0.853 |
| Magnesium | -0.0003 | 0.001 | -0.03 (-0.17, 0.11) | 0.664 | 0.853 |
| Mercury | 0.0008 | 0.002 | 0.08 (-0.24, 0.40) | 0.633 | 0.853 |

| Analyte | Coefficient | Std. Err | Increase (95% CI) | *p*-Value | BH-p |
|---|---|---|---|---|---|
| Potassium | -0.0005 | 0.001 | -0.05 (-0.23, 0.14) | 0.619 | 0.853 |
| Sulfate | -0.0005 | 0.001 | -0.05 (-0.25, 0.15) | 0.646 | 0.853 |
| Arsenic | -0.0006 | 0.002 | -0.06 (-0.48, 0.36) | 0.779 | 0.956 |
| Bicarbonate Alkalinity | 0.0001 | 0.002 | 0.01 (-0.46, 0.47) | 0.973 | 0.999 |
| Calcium | -0.0001 | 0.001 | -0.01 (-0.14, 0.12) | 0.927 | 0.999 |
| Cyanide | -0.0003 | 0.002 | -0.03 (-0.36, 0.30) | 0.859 | 0.999 |
| Manganese | -0.0003 | 0.002 | -0.03 (-0.51, 0.46) | 0.91 | 0.999 |
| Nitrate Nitrite | -0.0001 | 0.002 | -0.01 (-0.35, 0.32) | 0.933 | 0.999 |
| **Physical Parameters** | | | | | |
| Total Carbon | -0.001 | <0.0001 | -0.10 (-0.17, -0.02) | 0.008 | 0.05 |
| Color | 0.0016 | 0.001 | 0.16 (-0.08, 0.41) | 0.196 | 0.588 |
| Odor | -0.0012 | 0.001 | -0.12 (-0.30, 0.07) | 0.214 | 0.609 |
| pH | 0.0017 | 0.002 | 0.17 (-0.20, 0.54) | 0.363 | 0.784 |
| Aggressive Index | -0.0014 | 0.002 | -0.14 (-0.48, 0.20) | 0.412 | 0.807 |
| Hardness (Total as CaCO3) | -0.0006 | 0.001 | -0.06 (-0.23, 0.11) | 0.475 | 0.807 |

| Analyte | Coefficient | Std. Err | Increase (95% CI) | p-Value | BH-p |
|---|---|---|---|---|---|
| Total Dissolved Solids (TDS) | -0.0006 | 0.001 | -0.06 (-0.23, 0.11) | 0.478 | 0.807 |
| Conductivity, μmhos/cm at 25°C | -0.0006 | 0.001 | -0.06 (-0.23, 0.12) | 0.529 | 0.84 |
| Hydroxide as Calcium Carbonate | <0.0001 | 0.001 | -0.00 (-0.20, 0.20) | 0.972 | 0.999 |
| Turbidity | 0.0001 | 0.001 | 0.01 (-0.25, 0.28) | 0.923 | 0.999 |
| **Radionuclides** | | | | | |
| Radium 228 | 0.0072 | 0.002 | 0.72 (0.37, 1.08) | <0.001 | 0.002 |
| Combined Uranium | -0.0052 | 0.001 | -0.52 (-0.81, -0.23) | <0.001 | 0.006 |
| Gross Alpha Particle Activity | 0.0003 | 0.001 | 0.02 (-0.14, 0.19) | 0.758 | 0.952 |
| **SOCs** | | | | | |
| Foaming Agents/Surfactants | 0.0046 | 0.002 | 0.46 (0.14, 0.79) | 0.006 | 0.043 |
| Aldicarb Sulfone | 0.0029 | 0.002 | 0.29 (-0.12, 0.70) | 0.16 | 0.541 |
| Heptachlor | -0.0001 | 0.003 | -0.01 (-0.64, 0.63) | 0.982 | 0.999 |
| Heptachlor Epoxide | <0.0001 | 0.003 | 0.00 (-0.64, 0.64) | 0.999 | 0.999 |
| **VOCs** | | | | | |

| Analyte | Coefficient | Std. Err | Increase (95% CI) | *p*-Value | BH-p |
|---|---|---|---|---|---|
| 1,2-Dibromo-3-Chloropropane | -0.0056 | 0.002 | -0.55 (-0.89, -0.21) | 0.001 | 0.015 |
| p-Isopropyltoluene | 0.0035 | 0.002 | 0.35 (-0.10, 0.81) | 0.13 | 0.47 |
| 1,2,3-Trichloropropane | 0.0011 | 0.002 | 0.11 (-0.26, 0.47) | 0.564 | 0.847 |
| Trichlorofluoromethane | 0.0009 | 0.002 | 0.09 (-0.26, 0.43) | 0.617 | 0.853 |

**Table 2: Same specification as in Table 1, but with age-adjusted mortality rate as the outcome, and without age structure as adjustment covariates.**

| Analyte | Coefficient | Std. Err | Increase (95% CI) | P-Value | BH-p |
|---|---|---|---|---|---|
| **Disinfection byproducts** | | | | | |
| Chloroform | -0.0103 | 0.003 | -1.03 (-1.56, -0.49) | <0.001 | 0.004 |
| Dibromoacetic Acid | -0.0101 | 0.003 | -1.01 (-1.64, -0.37) | 0.002 | 0.015 |
| Dichloroacetic Acid | -0.0002 | 0.004 | -0.02 (-0.85, 0.82) | 0.97 | 0.97 |
| Trihalomethanes (TTHM) | -0.0063 | 0.003 | -0.63 (-1.16, -0.09) | 0.021 | 0.084 |

| Analyte | Coefficient | Std. Err | Increase (95% CI) | P-Value | BH-p |
|---|---|---|---|---|---|
| Ethylene Dibromide | -0.0022 | 0.004 | -0.22 (-0.91, 0.47) | 0.531 | 0.834 |
| Bromodichloromethane | -0.0061 | 0.003 | -0.61 (-1.20, -0.01) | 0.045 | 0.138 |
| Total Haloacetic Acids (HAA5) | 0.0004 | 0.003 | 0.04 (-0.56, 0.63) | 0.907 | 0.954 |
| Bromoform | 0.0004 | 0.002 | 0.04 (-0.42, 0.50) | 0.874 | 0.954 |
| Dibromochloromethane | -0.0032 | 0.003 | -0.32 (-0.95, 0.30) | 0.309 | 0.548 |
| Trichloroacetic Acid | 0.0088 | 0.005 | 0.89 (-0.06, 1.84) | 0.066 | 0.166 |
| **Inorganic Chemicals** | | | | | |
| Alkalinity (Carbonate) | 0.0039 | 0.001 | 0.39 (0.20, 0.59) | <0.001 | 0.002 |
| Perchlorate | 0.0072 | 0.003 | 0.72 (0.11, 1.34) | 0.02 | 0.084 |
| Iron | 0.0058 | 0.003 | 0.58 (-0.04, 1.21) | 0.065 | 0.166 |
| Total Alkalinity | -0.0065 | 0.003 | -0.64 (-1.22, -0.06) | 0.03 | 0.103 |
| Nitrate | -0.0074 | 0.004 | -0.73 (-1.55, 0.09) | 0.081 | 0.194 |
| Aluminum | -0.0002 | 0.003 | -0.02 (-0.60, 0.56) | 0.937 | 0.954 |
| Barium | 0.0046 | 0.002 | 0.46 (0.10, 0.82) | 0.011 | 0.057 |

| Analyte | Coefficient | Std. Err | Increase (95% CI) | P-Value | BH-p |
| --- | --- | --- | --- | --- | --- |
| Chromium Hex | -0.0009 | 0.003 | -0.09 (-0.65, 0.47) | 0.751 | 0.897 |
| Free Copper | 0.0018 | 0.001 | 0.18 (-0.11, 0.46) | 0.226 | 0.415 |
| Fluoride | 0.0023 | 0.003 | 0.23 (-0.41, 0.87) | 0.485 | 0.784 |
| Sodium | -0.0006 | 0.001 | -0.06 (-0.34, 0.22) | 0.667 | 0.854 |
| Chloride | -0.0012 | 0.001 | -0.12 (-0.24, 0.00) | 0.059 | 0.166 |
| Total Boron | 0.0003 | 0.003 | 0.03 (-0.52, 0.58) | 0.915 | 0.954 |
| Magnesium | -0.0008 | <0.001 | -0.08 (-0.15, -0.01) | 0.019 | 0.084 |
| Mercury | -0.001 | 0.002 | -0.10 (-0.42, 0.22) | 0.55 | 0.84 |
| Potassium | -0.0008 | 0.001 | -0.08 (-0.36, 0.20) | 0.57 | 0.847 |
| Sulfate | -0.0017 | 0.001 | -0.16 (-0.38, 0.05) | 0.138 | 0.292 |
| Arsenic | -0.0002 | 0.002 | -0.02 (-0.47, 0.43) | 0.927 | 0.954 |
| Alkalinity (Bicarbonate) | -0.0003 | 0.003 | -0.03 (-0.56, 0.50) | 0.907 | 0.954 |
| Calcium | -0.0004 | 0.001 | -0.04 (-0.19, 0.12) | 0.648 | 0.849 |
| Cyanide | -0.0006 | 0.002 | -0.06 (-0.39, 0.27) | 0.722 | 0.882 |

| Analyte | Coefficient | Std. Err | Increase (95% CI) | P-Value | BH-p |
| --- | --- | --- | --- | --- | --- |
| Manganese | -0.0014 | 0.003 | -0.14 (-0.72, 0.45) | 0.644 | 0.849 |
| Nitrate Nitrite | -0.0012 | <0.001 | -0.12 (-0.18, -0.05) | <0.001 | 0.006 |
| **Physical Parameters** | | | | | |
| Total Carbon | -0.0015 | <0.001 | -0.14 (-0.23, -0.06) | 0.002 | 0.014 |
| Color | 0.0026 | 0.002 | 0.26 (-0.09, 0.61) | 0.145 | 0.296 |
| Odor | -0.001 | 0.001 | -0.10 (-0.30, 0.10) | 0.324 | 0.557 |
| pH | 0.0032 | 0.002 | 0.32 (-0.13, 0.76) | 0.161 | 0.317 |
| Aggressive Index | -0.0027 | 0.002 | -0.27 (-0.62, 0.08) | 0.132 | 0.29 |
| Hardness (Total as CaCO3) | -0.0025 | 0.001 | -0.25 (-0.42, -0.08) | 0.004 | 0.024 |
| Total Dissolved Solids (TDS) | -0.0015 | 0.001 | -0.15 (-0.28, -0.01) | 0.03 | 0.103 |
| Conductivity, μmhos/cm at 25°C | -0.0018 | 0.001 | -0.18 (-0.30, -0.06) | 0.002 | 0.017 |
| Hydroxide as Calcium Carbonate | -0.0002 | 0.001 | -0.02 (-0.20, 0.17) | 0.865 | 0.954 |
| Turbidity | -0.0007 | 0.002 | -0.07 (-0.44, 0.30) | 0.707 | 0.882 |

| Analyte | Coefficient | Std. Err | Increase (95% CI) | P-Value | BH-p |
|---|---|---|---|---|---|
| **Radionuclides** | | | | | |
| Radium 228 | 0.0081 | 0.002 | 0.82 (0.46, 1.17) | <0.001 | <0.001 |
| Combined Uranium | -0.008 | 0.004 | -0.80 (-1.64, 0.05) | 0.066 | 0.166 |
| Gross Alpha Particle Activity | 0.0007 | 0.001 | 0.07 (-0.20, 0.34) | 0.606 | 0.849 |
| **SOCs** | | | | | |
| Foaming Agents/Surfactants | 0.0068 | 0.002 | 0.68 (0.22, 1.14) | 0.003 | 0.021 |
| Aldicarb Sulfone | 0.0027 | 0.002 | 0.27 (-0.16, 0.70) | 0.219 | 0.415 |
| Heptachlor | 0.0017 | 0.003 | 0.17 (-0.51, 0.85) | 0.628 | 0.849 |
| Heptachlor Epoxide | 0.0017 | 0.004 | 0.17 (-0.52, 0.87) | 0.624 | 0.849 |
| **VOCs** | | | | | |
| 1,2-Dibromo-3-Chloropropane | -0.0082 | 0.003 | -0.81 (-1.31, -0.32) | 0.001 | 0.014 |
| p-Isopropyltoluene | 0.004 | 0.002 | 0.41 (-0.06, 0.88) | 0.09 | 0.205 |
| 1,2,3-Trichloropropane | -0.0017 | 0.002 | -0.17 (-0.54, 0.21) | 0.379 | 0.632 |

| Analyte | Coefficient | Std. Err | Increase (95% CI) | P-Value | BH-p |
| --- | --- | --- | --- | --- | --- |
| Trichlorofluoromethane | 0.0038 | 0.002 | 0.38 (0.02, 0.74) | 0.038 | 0.122 |

Table 3: Same regression specifications as in Table 1, but with no covariate adjustments beyond zip code and year fixed effects.

| Analyte | Coefficient | Std. Err | Increase (95% CI) | P-Value | BH-p |
|---|---|---|---|---|---|
| **Disinfection byproducts** | | | | | |
| Chloroform | -0.0102 | 0.003 | -1.02 (-1.55, -0.48) | <0.001 | 0.003 |
| Dibromoacetic Acid | -0.0079 | 0.002 | -0.79 (-1.25, -0.32) | 0.001 | 0.011 |
| Dichloroacetic Acid | -0.0057 | 0.003 | -0.57 (-1.19, 0.06) | 0.077 | 0.279 |
| Trihalomethanes (TTHM) | -0.0042 | 0.003 | -0.42 (-0.92, 0.08) | 0.103 | 0.348 |
| Ethylene Dibromide | -0.003 | 0.003 | -0.30 (-0.98, 0.38) | 0.38 | 0.683 |
| Bromodichloromethane | -0.0024 | 0.003 | -0.24 (-0.77, 0.29) | 0.372 | 0.683 |
| Total Haloacetic Acids (HAA5) | -0.0025 | 0.003 | -0.25 (-0.78, 0.27) | 0.34 | 0.683 |
| Bromoform | 0.0033 | 0.002 | 0.33 (-0.09, 0.75) | 0.126 | 0.401 |
| Dibromochloromethane | 0.002 | 0.003 | 0.20 (-0.29, 0.70) | 0.423 | 0.692 |
| Trichloroacetic Acid | 0.0029 | 0.003 | 0.29 (-0.34, 0.92) | 0.362 | 0.683 |
| **Inorganic Chemicals** | | | | | |
| Alkalinity (Carbonate) | 0.0034 | 0.001 | 0.34 (0.18, 0.51) | <0.001 | <0.001 |

| Analyte | Coefficient | Std. Err | Increase (95% CI) | P-Value | BH-p |
|---|---|---|---|---|---|
| Perchlorate | 0.0075 | 0.003 | 0.76 (0.18, 1.34) | 0.011 | 0.057 |
| Iron | 0.0031 | 0.002 | 0.31 (-0.15, 0.78) | 0.189 | 0.538 |
| Total Alkalinity | -0.0057 | 0.003 | -0.57 (-1.11, -0.02) | 0.04 | 0.193 |
| Nitrate | -0.0077 | 0.004 | -0.76 (-1.49, -0.02) | 0.043 | 0.193 |
| Aluminum | 0.0021 | 0.003 | 0.21 (-0.38, 0.79) | 0.489 | 0.714 |
| Barium | 0.005 | 0.002 | 0.50 (0.19, 0.80) | 0.001 | 0.013 |
| Chromium Hex | 0.0027 | 0.002 | 0.27 (-0.21, 0.77) | 0.273 | 0.641 |
| Free Copper | 0.0008 | 0.001 | 0.08 (-0.20, 0.36) | 0.577 | 0.779 |
| Fluoride | 0.0061 | 0.003 | 0.62 (-0.00, 1.24) | 0.052 | 0.2 |
| Sodium | 0.0003 | 0.001 | 0.03 (-0.13, 0.19) | 0.722 | 0.817 |
| Chloride | 0.0006 | 0.001 | 0.06 (-0.08, 0.20) | 0.408 | 0.688 |
| Total Boron | -0.0015 | 0.003 | -0.15 (-0.72, 0.42) | 0.601 | 0.788 |
| Magnesium | -0.0004 | 0.001 | -0.04 (-0.18, 0.09) | 0.516 | 0.733 |
| Mercury | -0.0007 | 0.002 | -0.07 (-0.39, 0.26) | 0.685 | 0.817 |

| Analyte | Coefficient | Std. Err | Increase (95% CI) | P-Value | BH-p |
|---|---|---|---|---|---|
| Potassium | -0.0005 | 0.001 | -0.05 (-0.22, 0.12) | 0.575 | 0.779 |
| Sulfate | -0.0008 | 0.001 | -0.08 (-0.30, 0.14) | 0.463 | 0.714 |
| Arsenic | -0.0003 | 0.002 | -0.03 (-0.50, 0.45) | 0.901 | 0.941 |
| Alkalinity (Bicarbonate) | 0.0011 | 0.003 | 0.11 (-0.38, 0.61) | 0.657 | 0.807 |
| Calcium | -0.0001 | 0.001 | -0.01 (-0.17, 0.14) | 0.877 | 0.941 |
| Cyanide | -0.0022 | 0.002 | -0.22 (-0.56, 0.12) | 0.212 | 0.545 |
| Manganese | -0.0022 | 0.003 | -0.22 (-0.72, 0.29) | 0.399 | 0.688 |
| Nitrate Nitrite | -0.0022 | 0.002 | -0.22 (-0.59, 0.15) | 0.238 | 0.585 |
| **Physical Parameters** | | | | | |
| Total Carbon | -0.0009 | <0.001 | -0.09 (-0.15, -0.03) | 0.005 | 0.03 |
| Color | 0.0014 | 0.001 | 0.14 (-0.14, 0.42) | 0.322 | 0.683 |
| Odor | -0.0009 | 0.001 | -0.08 (-0.27, 0.10) | 0.364 | 0.683 |
| pH | 0.0032 | 0.002 | 0.32 (-0.11, 0.75) | 0.15 | 0.451 |
| Aggressive Index | -0.0006 | 0.002 | -0.06 (-0.39, 0.28) | 0.726 | 0.817 |

| Analyte | Coefficient | Std. Err | Increase (95% CI) | P-Value | BH-p |
| --- | --- | --- | --- | --- | --- |
| Hardness (Total as CaCO3) | -0.0009 | 0.001 | -0.09 (-0.27, 0.09) | 0.314 | 0.683 |
| Total Dissolved Solids (TDS) | -0.0005 | 0.001 | -0.05 (-0.19, 0.09) | 0.458 | 0.714 |
| Conductivity, μmhos/cm at 25°C | -0.0006 | 0.001 | -0.06 (-0.21, 0.10) | 0.486 | 0.714 |
| Hydroxide as Calcium Carbonate | 0.0001 | 0.001 | 0.01 (-0.16, 0.19) | 0.881 | 0.941 |
| Turbidity | 0.0001 | 0.001 | 0.01 (-0.25, 0.28) | 0.913 | 0.941 |
| **Radionuclides** | | | | | |
| Radium 228 | 0.0073 | 0.002 | 0.74 (0.38, 1.09) | <0.001 | <0.001 |
| Combined Uranium | -0.0046 | 0.002 | -0.46 (-0.78, -0.15) | 0.004 | 0.025 |
| Gross Alpha Particle Activity | 0.0004 | 0.001 | 0.04 (-0.12, 0.20) | 0.628 | 0.788 |
| **SOCs** | | | | | |
| Foaming Agents/Surfactants | 0.0066 | 0.002 | 0.66 (0.25, 1.08) | 0.002 | 0.013 |
| Aldicarb Sulfone | 0.0029 | 0.002 | 0.29 (-0.15, 0.73) | 0.202 | 0.545 |

| Analyte | Coefficient | Std. Err | Increase (95% CI) | P-Value | BH-p |
| --- | --- | --- | --- | --- | --- |
| Heptachlor | 0.0003 | 0.004 | 0.03 (-0.66, 0.72) | 0.941 | 0.941 |
| Heptachlor Epoxide | 0.0003 | 0.004 | 0.03 (-0.67, 0.73) | 0.937 | 0.941 |
| **VOCs** | | | | | |
| 1,2-Dibromo-3-Chloropropane | -0.008 | 0.002 | -0.80 (-1.14, -0.45) | <0.001 | <0.001 |
| p-Isopropyltoluene | 0.0051 | 0.003 | 0.51 (0.01, 1.01) | 0.046 | 0.193 |
| 1,2,3-Trichloropropane | 0.0008 | 0.002 | 0.08 (-0.34, 0.49) | 0.721 | 0.817 |
| Trichlorofluoromethane | -0.0009 | 0.002 | -0.09 (-0.45, 0.26) | 0.615 | 0.788 |

Table 4: Same regression specification as Table 1, but with no covariate adjustment for water source.

| Analyte | Coefficient | Std. Err | Increase (95% CI) | P-Value | BH-p |
|---|---|---|---|---|---|
| **Disinfection byproducts** | | | | | |
| Chloroform | -0.0099 | 0.003 | -0.98 (-1.50, -0.46) | <0.001 | 0.004 |
| Dibromoacetic Acid | -0.0067 | 0.002 | -0.66 (-1.08, -0.25) | 0.002 | 0.014 |
| Dichloroacetic Acid | -0.0062 | 0.003 | -0.62 (-1.23, -0.01) | 0.047 | 0.21 |
| Trihalomethanes (TTHM) | -0.0039 | 0.002 | -0.39 (-0.87, 0.09) | 0.108 | 0.418 |
| Ethylene Dibromide | -0.0041 | 0.003 | -0.40 (-1.02, 0.21) | 0.199 | 0.597 |
| Bromodichloromethane | -0.0024 | 0.003 | -0.24 (-0.74, 0.26) | 0.346 | 0.779 |
| Total Haloacetic Acids (HAA5) | -0.0025 | 0.002 | -0.25 (-0.73, 0.23) | 0.305 | 0.779 |
| Bromoform | 0.0014 | 0.002 | 0.14 (-0.24, 0.52) | 0.47 | 0.805 |
| Dibromochloromethane | 0.0017 | 0.002 | 0.17 (-0.27, 0.61) | 0.449 | 0.805 |
| Trichloroacetic Acid | 0.0001 | 0.003 | 0.01 (-0.59, 0.61) | 0.981 | 0.995 |
| **Inorganic Chemicals** | | | | | |
| Alkalinity (Carbonate) | 0.0028 | 0.001 | 0.28 (0.15, 0.41) | <0.001 | 0.001 |

| Analyte | Coefficient | Std. Err | Increase (95% CI) | P-Value | BH-p |
|---|---|---|---|---|---|
| Perchlorate | 0.0074 | 0.003 | 0.74 (0.19, 1.30) | 0.008 | 0.051 |
| Iron | 0.0045 | 0.002 | 0.45 (0.04, 0.87) | 0.033 | 0.177 |
| Total Alkalinity | -0.0049 | 0.003 | -0.49 (-0.98, 0.00) | 0.052 | 0.217 |
| Nitrate | -0.0073 | 0.004 | -0.72 (-1.43, -0.01) | 0.046 | 0.21 |
| Aluminum | 0.0027 | 0.003 | 0.27 (-0.28, 0.83) | 0.333 | 0.779 |
| Barium | 0.0014 | 0.001 | 0.14 (-0.15, 0.42) | 0.346 | 0.779 |
| Chromium Hex | 0.0022 | 0.002 | 0.22 (-0.23, 0.68) | 0.343 | 0.779 |
| Free Copper | 0.0011 | 0.001 | 0.11 (-0.18, 0.40) | 0.447 | 0.805 |
| Fluoride | 0.0022 | 0.003 | 0.22 (-0.32, 0.77) | 0.421 | 0.805 |
| Sodium | 0.0005 | 0.001 | 0.05 (-0.10, 0.20) | 0.518 | 0.834 |
| Chloride | 0.0005 | 0.001 | 0.05 (-0.11, 0.21) | 0.549 | 0.84 |
| Total Boron | -0.0013 | 0.003 | -0.13 (-0.71, 0.46) | 0.66 | 0.849 |
| Magnesium | -0.0003 | 0.001 | -0.03 (-0.17, 0.11) | 0.66 | 0.849 |
| Mercury | 0.0008 | 0.002 | 0.08 (-0.24, 0.40) | 0.615 | 0.849 |

| Analyte | Coefficient | Std. Err | Increase (95% CI) | P-Value | BH-p |
|---|---|---|---|---|---|
| Potassium | -0.0005 | 0.001 | -0.05 (-0.23, 0.14) | 0.619 | 0.849 |
| Sulfate | -0.0005 | 0.001 | -0.05 (-0.25, 0.15) | 0.646 | 0.849 |
| Arsenic | -0.0006 | 0.002 | -0.06 (-0.48, 0.36) | 0.775 | 0.951 |
| Alkalinity (Bicarbonate) | <0.0001 | 0.002 | 0.00 (-0.46, 0.47) | 0.986 | 0.995 |
| Calcium | -0.0001 | 0.001 | -0.01 (-0.14, 0.12) | 0.926 | 0.995 |
| Cyanide | -0.0002 | 0.002 | -0.02 (-0.36, 0.31) | 0.884 | 0.995 |
| Manganese | -0.0003 | 0.002 | -0.03 (-0.51, 0.46) | 0.909 | 0.995 |
| Nitrate Nitrite | -0.0001 | 0.002 | -0.01 (-0.35, 0.32) | 0.934 | 0.995 |
| **Physical Parameters** | | | | | |
| Total Carbon | -0.001 | <0.001 | -0.10 (-0.17, -0.02) | 0.008 | 0.051 |
| Color | 0.0016 | 0.001 | 0.16 (-0.09, 0.41) | 0.199 | 0.597 |
| Odor | -0.0012 | 0.001 | -0.12 (-0.30, 0.07) | 0.215 | 0.61 |
| pH | 0.0017 | 0.002 | 0.17 (-0.20, 0.54) | 0.365 | 0.788 |
| Aggressive Index | -0.0014 | 0.002 | -0.14 (-0.48, 0.20) | 0.405 | 0.805 |

| Analyte | Coefficient | Std. Err | Increase (95% CI) | P-Value | BH-p |
|---|---|---|---|---|---|
| Hardness (Total as CaCO3) | -0.0006 | 0.001 | -0.06 (-0.23, 0.11) | 0.467 | 0.805 |
| Total Dissolved Solids (TDS) | -0.0006 | 0.001 | -0.06 (-0.23, 0.11) | 0.477 | 0.805 |
| Conductivity, μmhos/cm at 25°C | -0.0006 | 0.001 | -0.06 (-0.23, 0.12) | 0.525 | 0.834 |
| Hydroxide as Calcium Carbonate | <0.0001 | 0.001 | -0.00 (-0.20, 0.20) | 0.973 | 0.995 |
| Turbidity | 0.0001 | 0.001 | 0.01 (-0.25, 0.28) | 0.922 | 0.995 |
| **Radionuclides** | | | | | |
| Radium 228 | 0.0072 | 0.002 | 0.72 (0.37, 1.08) | <0.001 | 0.002 |
| Combined Uranium | -0.0052 | 0.001 | -0.52 (-0.81, -0.23) | <0.001 | 0.006 |
| Gross Alpha Particle Activity | 0.0003 | 0.001 | 0.02 (-0.14, 0.19) | 0.76 | 0.951 |
| **SOCs** | | | | | |
| Foaming Agents/Surfactants | 0.0046 | 0.002 | 0.46 (0.13, 0.79) | 0.006 | 0.05 |
| Aldicarb Sulfone | 0.0029 | 0.002 | 0.29 (-0.12, 0.70) | 0.16 | 0.539 |

| Analyte | Coefficient | Std. Err | Increase (95% CI) | P-Value | BH-p |
|---|---|---|---|---|---|
| Heptachlor | -0.0001 | 0.003 | -0.00 (-0.64, 0.63) | 0.987 | 0.995 |
| Heptachlor Epoxide | <0.0001 | 0.003 | 0.00 (-0.64, 0.64) | 0.995 | 0.995 |
| **VOCs** | | | | | |
| 1,2-Dibromo-3-Chloropropane | -0.0056 | 0.002 | -0.56 (-0.89, -0.22) | 0.001 | 0.014 |
| p-Isopropyltoluene | 0.0035 | 0.002 | 0.35 (-0.10, 0.81) | 0.131 | 0.47 |
| 1,2,3-Trichloropropane | 0.0011 | 0.002 | 0.11 (-0.25, 0.47) | 0.56 | 0.84 |
| Trichlorofluoromethane | 0.0009 | 0.002 | 0.09 (-0.25, 0.44) | 0.597 | 0.849 |

**Table 5: Same regression specification as in Table 1, but with only adjusting for age structure.**

| Analyte | Coefficient | Std. Err | Increase (95% CI) | P-Value | BH-p |
|---|---|---|---|---|---|
| **Disinfection byproducts** | | | | | |
| Chloroform | -0.0097 | 0.003 | -0.96 (-1.48, -0.44) | <0.001 | 0.003 |
| Dibromoacetic Acid | -0.0083 | 0.002 | -0.82 (-1.25, -0.39) | <0.001 | 0.003 |
| Dichloroacetic Acid | -0.0045 | 0.003 | -0.45 (-1.06, 0.16) | 0.15 | 0.413 |
| Trihalomethanes (TTHM) | -0.0036 | 0.003 | -0.36 (-0.84, 0.13) | 0.153 | 0.413 |
| Ethylene Dibromide | -0.004 | 0.003 | -0.40 (-1.04, 0.25) | 0.229 | 0.547 |
| Bromodichloromethane | -0.0019 | 0.003 | -0.19 (-0.70, 0.32) | 0.467 | 0.7 |
| Total Haloacetic Acids (HAA5) | -0.0014 | 0.002 | -0.14 (-0.62, 0.35) | 0.586 | 0.811 |
| Bromoform | 0.0021 | 0.002 | 0.21 (-0.19, 0.60) | 0.3 | 0.579 |
| Dibromochloromethane | 0.0015 | 0.002 | 0.15 (-0.32, 0.62) | 0.527 | 0.749 |
| Trichloroacetic Acid | 0.0036 | 0.003 | 0.36 (-0.24, 0.97) | 0.238 | 0.547 |
| **Inorganic Chemicals** | | | | | |
| Alkalinity (Carbonate) | 0.0035 | 0.001 | 0.35 (0.18, 0.51) | <0.001 | <0.001 |

| Analyte | Coefficient | Std. Err | Increase (95% CI) | P-Value | BH-p |
| --- | --- | --- | --- | --- | --- |
| Perchlorate | 0.0077 | 0.003 | 0.78 (0.21, 1.34) | 0.007 | 0.037 |
| Iron | 0.0041 | 0.002 | 0.41 (-0.04, 0.86) | 0.075 | 0.313 |
| Total Alkalinity | -0.0053 | 0.003 | -0.53 (-1.04, -0.01) | 0.047 | 0.213 |
| Nitrate | -0.0082 | 0.004 | -0.82 (-1.53, -0.10) | 0.026 | 0.126 |
| Aluminum | 0.0025 | 0.003 | 0.25 (-0.32, 0.82) | 0.391 | 0.64 |
| Barium | 0.004 | 0.001 | 0.41 (0.11, 0.70) | 0.007 | 0.037 |
| Chromium Hex | 0.001 | 0.002 | 0.10 (-0.38, 0.57) | 0.691 | 0.904 |
| Free Copper | 0.0011 | 0.001 | 0.11 (-0.17, 0.39) | 0.432 | 0.666 |
| Fluoride | 0.0046 | 0.003 | 0.46 (-0.11, 1.04) | 0.114 | 0.395 |
| Sodium | 0.0001 | 0.001 | 0.01 (-0.16, 0.17) | 0.939 | 0.985 |
| Chloride | -0.0001 | 0.001 | -0.01 (-0.19, 0.17) | 0.913 | 0.985 |
| Total Boron | -0.0003 | 0.003 | -0.03 (-0.64, 0.58) | 0.913 | 0.985 |
| Magnesium | -0.001 | 0.001 | -0.10 (-0.30, 0.10) | 0.346 | 0.602 |
| Mercury | -0.0005 | 0.002 | -0.05 (-0.37, 0.26) | 0.737 | 0.904 |

| Analyte | Coefficient | Std. Err | Increase (95% CI) | P-Value | BH-p |
|---|---|---|---|---|---|
| Potassium | -0.0008 | 0.001 | -0.08 (-0.27, 0.12) | 0.426 | 0.666 |
| Sulfate | -0.0012 | 0.001 | -0.12 (-0.36, 0.13) | 0.343 | 0.602 |
| Arsenic | -0.0001 | 0.002 | -0.01 (-0.47, 0.45) | 0.966 | 0.985 |
| Alkalinity (Bicarbonate) | 0.0002 | 0.002 | 0.02 (-0.46, 0.50) | 0.933 | 0.985 |
| Calcium | -0.0003 | 0.001 | -0.03 (-0.14, 0.09) | 0.652 | 0.88 |
| Cyanide | -0.0015 | 0.002 | -0.15 (-0.48, 0.19) | 0.386 | 0.64 |
| Manganese | -0.0017 | 0.003 | -0.17 (-0.69, 0.35) | 0.52 | 0.749 |
| Nitrate Nitrite | -0.0026 | 0.002 | -0.26 (-0.61, 0.10) | 0.152 | 0.413 |
| **Physical Parameters** | | | | | |
| Total Carbon | -0.0009 | <0.001 | -0.09 (-0.16, -0.03) | 0.003 | 0.023 |
| Color | 0.0021 | 0.001 | 0.21 (-0.05, 0.47) | 0.117 | 0.395 |
| Odor | -0.0011 | 0.001 | -0.11 (-0.29, 0.08) | 0.262 | 0.565 |
| pH | 0.0025 | 0.002 | 0.26 (-0.14, 0.66) | 0.211 | 0.542 |
| Aggressive Index | -0.0017 | 0.002 | -0.17 (-0.51, 0.17) | 0.315 | 0.587 |

| Analyte | Coefficient | Std. Err | Increase (95% CI) | P-Value | BH-p |
|---|---|---|---|---|---|
| Hardness (Total as CaCO3) | -0.0014 | 0.001 | -0.14 (-0.36, 0.09) | 0.243 | 0.547 |
| Total Dissolved Solids (TDS) | -0.0012 | 0.001 | -0.12 (-0.33, 0.10) | 0.282 | 0.579 |
| Conductivity, μmhos/cm at 25°C | -0.0012 | 0.001 | -0.12 (-0.34, 0.10) | 0.295 | 0.579 |
| Hydroxide as Calcium Carbonate | 0.0002 | 0.001 | 0.02 (-0.19, 0.23) | 0.83 | 0.985 |
| Turbidity | <0.0001 | 0.001 | 0.00 (-0.26, 0.27) | 0.995 | 0.995 |
| **Radionuclides** | | | | | |
| Radium 228 | 0.0073 | 0.002 | 0.73 (0.38, 1.08) | <0.001 | <0.001 |
| Combined Uranium | -0.0052 | 0.002 | -0.52 (-0.83, -0.20) | 0.001 | 0.01 |
| Gross Alpha Particle Activity | 0.0003 | 0.001 | 0.03 (-0.13, 0.19) | 0.726 | 0.904 |
| **SOCs** | | | | | |
| Foaming Agents/Surfactants | 0.0061 | 0.002 | 0.62 (0.24, 0.99) | 0.001 | 0.01 |
| Aldicarb Sulfone | 0.0031 | 0.002 | 0.31 (-0.10, 0.73) | 0.141 | 0.413 |

| Analyte | Coefficient | Std. Err | Increase (95% CI) | P-Value | BH-p |
|---|---|---|---|---|---|
| Heptachlor | -0.0003 | 0.003 | -0.03 (-0.68, 0.63) | 0.929 | 0.985 |
| Heptachlor Epoxide | -0.0002 | 0.003 | -0.02 (-0.68, 0.64) | 0.943 | 0.985 |
| **VOCs** | | | | | |
| 1,2-Dibromo-3-Chloropropane | -0.007 | 0.002 | -0.69 (-1.03, -0.35) | <0.001 | 0.001 |
| p-Isopropyltoluene | 0.0041 | 0.002 | 0.41 (-0.05, 0.87) | 0.082 | 0.316 |
| 1,2,3-Trichloropropane | 0.0001 | 0.002 | 0.01 (-0.35, 0.37) | 0.966 | 0.985 |
| Trichlorofluoromethane | 0.0006 | 0.002 | 0.06 (-0.28, 0.40) | 0.717 | 0.904 |